Saturation-dependence of dispersion in porous media

B. Ghanbarian-Alavijeh

T. E. Skinner

A. G. Hunt

**Abstract**

In this study, we develop a saturation-dependent treatment of dispersion in porous media using concepts from critical path analysis, cluster statistics of percolation, and fractal scaling of percolation clusters. We calculate spatial solute distributions as a function of time and calculate arrival time distributions as a function of system size. Our previous results correctly predict the range of observed dispersivity values over ten orders of magnitude in experimental length scale, but that theory contains no explicit dependence on porosity or relative saturation. This omission complicates comparisons with experimental results for dispersion, which are often conducted at saturation less than 1. We now make specific comparisons of our predictions for the arrival time distribution with experiments on a single column over a range of saturations. This comparison suggests that the most important predictor of such distributions as a function of saturation is not the value of the saturation per se, but the applicability of either random or invasion percolation models, depending on experimental conditions.

**Introduction**

*Existing models for solute transport in porous media*

Predicting dispersion in porous media is relevant to a wide range of problems in applied physics, such as remediation and monitoring of toxic wastes in groundwater, cellular mitosis, blood perfusion in the brain, chromatography, filtration, secondary oil recovery, catalysis, and the behavior of packed bed reactors [1], degradation of building materials [2,3], tissue physiology [4], migration and epidemiology [5,6], as well as heat dispersion in foams [7] and the internal dynamics of the atom [8]. Dispersion is a process in which particles move apart due to molecular diffusion and heterogeneities in a vector flow field [1]. The typical approach to modeling dispersion, particularly in the groundwater flow focused on here, is to use the partial differential equation referred to most frequently as the advection-dispersion equation (ADE) from continuum mechanics (in its traditional form) [9], but known also as the convection-dispersion equation (CDE). Advection is a process by which fluids flowing through the medium carry along solutes passively; thus, the variability of local flow velocities leads to a spreading, or dispersion, in a solute plume. The ADE treats advection using a scalar product of the local fluid velocity and the gradient of the concentration while addressing diffusion (or hydrodynamic dispersion) using a dispersion constant times the Laplacian operator.

For a variety of reasons, however, the ADE is not a satisfactory model for solute transport in groundwater [10-12]. In particular, it predicts a normal (Gaussian) distribution for solute arrival times, whereas experimental results, both in porous and fractured media [11], typically conform more nearly to power-law distribution tails at long times. While such failures may not be surprising in fractured media (which may be fractal), the reason for such a failure in an ostensibly homogeneous sand column are not as clear. Moreover, the ADE underestimates solute transport at both short and long times as pointed out by Cortis and Berkowitz [11]. The observed long-tailed arrival time distributions are spatially invariant, i.e., they have the same shape for long transport distances as for short distances [13, 14]. This shape invariance produces solute spreading (the variance of the distribution) proportional to the square of the travel distance, in contrast to Gaussian distributions, where the variance is linearly proportional to the travel time (and thus distance) [13, 14]. The relevance of the long-time tail to groundwater remediation, for example, is that the time required for toxic waste concentrations to fall below an acceptable level may be much greater than for a Gaussian distribution of arrival times [15]. Essentially identical behavior has been identified in the "dispersive transport" of photo-excited electrons in amorphous semiconductors and polymers [13,14].

The ubiquity of this long-tailed particle arrival time distribution makes the solution to this problem for transport in porous media and dispersive electronic transport in disordered materials widely relevant [10-14]. Many alternative frameworks for understanding transport have been proposed, including the Fractional Advection-Dispersion Equation (FADE) [15-22], and the continuous time random walk (CTRW) [10,11,13,14,23-25]. Since the FADE still underestimates solute arrivals at short times [11] and the dispersion coefficient of the FADE can still be scale-dependent [17], its relevance may be chiefly in the field of mathematics rather than to actual solute transport problems. Although the generality of the CTRW makes it an attractive and promising choice for modelers, values for the specific parameters which govern the truncated power-law arrival time distribution at long times are unknown, which hinders its use for specific predictions. One aspect of the CTRW that speaks strongly in its favor is that it may be modified in a straightforward fashion to account for non-conservative solutes [26], which may be quite difficult for the treatment considered here.

In order to address the failures of the conventional ADE, the field of groundwater hydrology has considered stochastic differential equations (stochastic subsurface hydrology [27-35]), in which the parameters of the equation are allowed to vary stochastically. While such techniques can, under ideal circumstances (constantly evolving heterogeneity with increasing spatial scale), be made to generate a variance of the distribution that increases faster than linearly [36], they do not generate the required spatially invariant long tails in the arrival time distribution. Furthermore, the existing predictions [37] of the stochastic theory regarding the variance have also been shown [12] to underestimate observed values in micromodel experiments by roughly 4 orders of magnitude.

We have documented numerous successes of our alternative theory for solute transport (e.g., [12, 38-40]), but solute distributions in our existing publications contain no explicit dependence on porosity or relative saturation. This omission complicates comparisons with experimental results for dispersion, which are often conducted at saturation less than 1. Since mixing from diffusion was addressed for *saturated conditions* [39] and found to be relatively unimportant except for the smallest length scales, diffusion was neglected in the present extension to unsaturated conditions. Solute dispersion appears to be more significantly affected by spatially variable solute advection [12]. This variability is a product of the variability of the local values of the hydraulic conductivity (or conductances) and their correlations. The importance of the correlations lies in their influence on the connectivity of the local regions of higher conductance, a topic which may be more suited to discussion in terms of percolation theory.

In the present paper, we therefore consider the saturation-dependence of the percolation model and also correct a minor omission in the original model. We demonstrate that resulting calculation errors were minor, producing no discernible change in the original results. We develop analytical expressions with explicit reference to saturation from which solute distributions and their most important moments can be derived numerically. Consequently, we can now make direct comparisons with additional experiments. Finally, we generalize the model of the medium itself in order to be able to treat a wider range of porous media.

**Theory**

We first provide a brief overview of the previous theory applicable to saturated conditions, providing the context for the modifications required to include saturation dependence in the model. To generate a prediction for dispersion, we choose a model of the porous medium and use it to develop the appropriate statistical representation of the impedance network. We then use the framework of percolation theory (including critical path analysis) to treat the transport paths through the network.

*Percolation theoretic approach*

Close links between the spatial variability of the hydraulic conductivity, $K$, and solute dispersion have been noted in the stochastic theory of flow and transport in porous media [28-32]. Issues of flow channeling or preferential flow become more important for more heterogeneous media, increasing the likelihood that transport will be strongly influenced by these optimal flow paths. The links between flow and transport are emphasized further when the physics of transport is examined within the context of percolation theory. We have derived a highly successful method for predicting dispersion in porous media that predicts [12,38-41]:

- The observed distribution of dispersivity values over ten orders of magnitude of length scale (consistent with over 2200 experiments) [12] and the dependences of individual experimental results for the dispersivity on length scale and heterogeneity [12],

- The temporal evolution of the variance in a series of Borden aquifer experiments (without adjustable parameters, or even detailed knowledge of the subsurface) [40],

- The observed distribution [42] of arrival times in two-dimensional simulations at the percolation threshold [38], giving also the predominant behavior of solute arrival time tails in fracture flow [43,44],

- The systematic variation [12] of the powers in the scaling of system crossing times with system length in dispersive transport with dimensionality of the random component of the flow variability in (3-D) amorphous semiconductors (ca. 1.9) [45-47] and (2-D) polymers (ca. 1.64) [48-50].

- The observed variability of these powers, if related to the influence of finite-size corrections (ca. 10%) [12,41].

Note that powers of the time-length scaling greater than one imply a reduction in mean particle velocity with distance.

In our derivation, two hitherto unrecognized fundamental connections between solute dispersion and the hydraulic conductivity were established: 1) the predicted diminution of mean solute velocity with increasing distances [12] corresponds to a diminution of *K* with increasing length scale (verified by direct comparison with numerical results [51]), 2) the influence of the experimenter on the fundamental length scale controlling the dispersivity, as we determined in [40], is paralleled by the role of the support volume in determining the range of the correlations in the hydraulic conductivity semi-variogram [36,37,52]. In the case of anisotropy, an observed *increase* of *K* with increasing length scale [36,37] is more properly interpreted as a shape, rather than as a scale, effect [53,54]. Thus, with increasing system size, the experimental volume (which effectively is highly non-equidimensional) accommodates a cross-over in flow path structure from one-dimensional to three-dimensional, with corresponding reduction in the critical volume fraction for percolation and associated increase in *K*. These relationships are best understood within the framework of critical path analysis from percolation theory [55], which clarifies that, at large length scales, important flow and dominant solute transport paths are controlled by the topology of percolation.

*Basis for the saturation-dependence of the percolation model*

The explicit saturation dependence of the dispersion is calculated assuming that pores are filled completely either with water or air, with the distinction dependent on the interfacial tension of the two phases as represented by the Young-Laplace equation. This particular approximation is common in soils, but is not necessarily justifiable and is not typically used in other areas of research in porous media. Thus, if warranted, later improvements to our calculations should involve the relaxation of this assumption. While the applicability of our calculations is intended to be restricted only to saturations for which the water phase is continuous, the conditions determining which experimental data were suitable for comparison did restrict severely the range of saturation values that actually appear in our assessment.

As the saturation of a porous medium is reduced, it is known that two percolation transitions occur [55,60]: the first occurs when the invading non-wetting fluid first percolates, and the second occurs when the defending, wetting, fluid first ceases to percolate. When the wetting fluid is water, this moisture content is typically referred to as the residual saturation. When the wetting fluid is reintroduced, two analogous percolation transitions occur [55,60]. When the non-wetting fluid is air, air entrapment is possible at the second percolation transition, since the air phase becomes discontinuous and a connected path for air flow no longer exists. How do these changes affect solute transport? The latter eventuality appears to be the most important single consideration.

But consider now also the varying physical inputs to the hydraulic conductivity as a function of saturation. The hydraulic conductivity drops rapidly with diminishing saturation, but for different reasons at high and low moisture contents [61]. At high saturations, the hydraulic conductivity is dominated by pore distribution effects, for which the *optimal* flow paths are governed by structural constraints, whereas at the critical limit for saturation, the topology of the *allowed* flow paths is governed by percolation theory. The theoretical changes that we have introduced, which accommodate the changes in flow constraints as a function of saturation, are able to distinguish the cases of structural constraints on flow (low saturation) and optimization of flow paths in flow heterogeneity by critical path analysis (high saturation). In the latter case we account for all solute transport paths, but use the structure and language of percolation to classify them.

The local variation in transport resistance due to variability in pore size or medium characteristics can be represented using a Master Equation (see e.g., Scher et al. [10] and [57]). Such a Master Equation can be transformed to an impedance network representation [51], in which the impedance distribution defines the local resistance to flow [51]. Our existing development [12,38-41] starts with a combination of critical path analysis and the cluster statistics of percolation theory to predict the probability density function, $W_p(g|x)$ that a given system of length $x$ can be spanned by an interconnected cluster of conductances with minimum local conductance value, $g$, and cluster length equal to the system length $x$ (ie, the clusters span the system). Then we use the theory for the topology of such clusters to predict the dependence of the cluster crossing time, $t(g,x)$ on cluster length, $x$, as well as minimal conductance value, $g$. In this theory the dominant parameter is the fractal dimensionality of the percolation backbone [59]. From fundamental probability theory we can then predict the distribution of solute arrival times in terms of $W_p(g|x)$ and $t(g,x)$. Although this theory was developed to address solute transport in a medium where flow heterogeneity is due to variability in the local hydraulic conductances, we are now able to treat heterogeneity which is purely structural. The structural phase transition can be addressed by choosing parameters that yield a narrow distribution of local conductances for values of the saturation near the percolation threshold.

*Model*

In prior publications, we have adopted the Rieu and Sposito (RS) truncated random fractal model [62] of a porous medium. While this is the simplest realistic model that could be applied, here we prefer generalizing the existing treatment to accommodate a somewhat wider variety of media. Our new treatment is a generalization of the RS model that yields the Tyler and Wheatcraft [63] fractal model and RS [62] models as its limiting cases. The procedure here will be to present the relevant arguments, then to present the new results alongside the older ones.

The chosen network representation here is based on a two phase (pore and solid) model. As in past publications, we found it convenient to use a continuous distribution of pore sizes (in terms of a probability density function, or pdf) rather than using discrete fractal treatments [51]. In this treatment, the pdf describing the pore sizes is written as,

$$W(r) = kr^{-1-D}, \quad r_{\min} < r < r_{\max} \tag{1}$$

where $k$ is a constant coefficient (a normalization factor), $r$ is the pore radius, $D$ is the fractal dimensionality of the pore space, and $r_{\min}$ and $r_{\max}$ are the lower and upper limits of the (truncated) fractal distribution.

If one takes the integral of Eq. (1) from $r_{min}$ to $r_{max}$ and sets it equal to 1, the constant coefficient $k$ would be:

$$k = \frac{D}{r_{min}^{-D} - r_{max}^{-D}} \tag{2}$$

In natural porous media for which $r_{min} \ll r_{max}$, Eq. (2) reduces to:

$$k = Dr_{min}^{D} \tag{3}$$

Only for this simplification does our treatment yield the Tyler and Wheatcraft model [63]. However, we note that for the purpose of examining (primarily artificial) porous media with limited ranges of pore sizes, this approximation cannot be used. The simplification in Eq. (3) may also compromise the accuracy of predictions for $D$ values much smaller than 2.

Defining porosity as the integrated pore volume between $r_{min}$ and $r_{max}$, weighted by the pdf at each value of $r$, using Eqs. (1) and (3) gives

$$\phi = \int_{r_{min}}^{r_{max}} Br'^3 W(r')dr' = \frac{BDr_{min}^D}{3-D}\left[r_{max}^{3-D} - r_{min}^{3-D}\right] \tag{4}$$

where $B$ is a pore shape factor.

The water content as a function of $r$ can also be defined as

$$\theta = \int_{r_{min}}^{r} Br'^3 W(r')dr' = \frac{BDr_{min}^D}{3-D}\left[r^{3-D} - r_{min}^{3-D}\right] \tag{5}$$

Combining Eqs. (4) and (5) with the Young-Laplace equation (specifically, $h = A/r$ in which $A$ is a constant and $h$ is tension head) gives the soil water retention curve

$$\theta = \phi - \beta\left[1 - \left(\frac{h}{h_{min}}\right)^{D-3}\right], \quad h_{min} < h < h_{max} \tag{6}$$

in which $\beta = \frac{\phi r_{max}^{3-D}}{r_{max}^{3-D} - r_{min}^{3-D}}$. The bounds on the tension are $h_{min} = A/r_{max}$ and $h_{max} = A/r_{min}$.

Our general model for the soil water retention curve in Eq. (6) reduces to the Tyler and Wheatcraft [63] and Rieu and Sposito [62] models for $\beta$ equal to $\phi$ and 1, respectively. This increased generality means that it is possible to address a wider range of media. The chief advantage here will be that we can, for a wider range of natural media, make specific predictions for the dispersion that are based on specific parameters obtained from experimental soil water retention curves. Eq. (6) is consistent with a model developed by Perrier et al. [64] in which $\beta = V_0/V_T$, where $V_0$ is the upper bound on the total pore volume as $r_{min}$ approaches 0 and $V_T$ is the total volume of a soil sample. The parameters of the generalized soil water retention model (Eq. 6) can be extracted from an experimental soil water retention curve (tension head $h$ as a function of water content $\theta$) [65].

The soil water retention curve (SWRC), Eq. (6), obtained from the pore-size distribution, assumes that pores with radius greater than a given value, $r$, are filled with air, while smaller pores are filled with water. The Young-Laplace equation gives $r = A/h$ as the inverse of the tension head, $h$, and $A$ is a parameter that depends on the wettability of the surface. As described in what follows, the pore-size distribution is then used to find the distribution of conductance values, $W(g)$, to put into the random impedance network [51]. An important purpose of including the development of the SWRC for the general model (Eq. 6) is to give a result that allows extraction from experiment of all the relevant parameters for the prediction of dispersion. Using an experimental SWRC to find the relevant values of $\beta$, $D$, $\theta_t$ cannot be expected to give valid results, however, if the experimental SWRC does not conform to the general model; thus we do not yet have a universal algorithm for predicting dispersion.

From the Poiseiulle flow equation, one can generate a conductance [51,66],

$$g \propto \frac{r^4}{l} \propto r^3, \tag{7}$$

assuming the pore radius $r$ and pore length $l$ are proportional to each other, in order to conform to the requirements of self-similarity, consistent with use of a fractal model. The pdf for the local conductance distribution, $W(g)$, is found from $W(g)\,dg = W(r)\,dr$ using Eq. (7) for $g(r)$.

Quantities in percolation theory can be expressed in terms of the percolation variable $p$, which, in continuum percolation, is a fractional volume. In the case of variable saturation, when one is interested in the connectivity of the wetting phase, $p$ is the fractional moisture content. If one is interested in defining an optimal flow path, $p$ must be related to a conductance distribution. One can relate [51,55] $p$ related to the cumulative probability distribution $W'(g)$ of conductances less than a given value, $g$, as follows:

$$p = 1 - W'(g) = \int_g^\infty W(g')\,dg' \tag{8}$$

The special case

$$p_c = 1 - W'(g_c) = \int_{g_c}^\infty W(g')\,dg' , \tag{9}$$

the percolating value of $p$, defines the interconnected subnetwork of the system with the largest possible value of the minimum conductance [51,55], $g_c$.

In this case then, $g_c$ is referred to as the critical conductance because it corresponds to critical percolation. In continuum percolation problems the variables $p$ and $p_c$ correspond to a volume fraction, $V$, and a critical volume fraction, $V_c$, respectively. We choose a continuum percolation representation since we are interested in percolating fractions of the pore space, and since we thereby sidestep the need to classify this space as pore throat or pore body, as well as the need to understand local coordination numbers of pores. The drawback is that we cannot, a priori, make an estimate of the value of the critical volume fraction for percolation. However, this is not such a practical difficulty in the present case, as we will be considering soil water retention curves, and $V_c$ may typically be identified with the minimum water content attained, $\theta_t$ [66]. Still, in the context of choosing values of exponents we will have to revisit the question of what kind of percolation problem drainage corresponds to.

The pore-size distribution of Eq. (1) is a power-law distribution, which also gives a power-law distribution for $W(g)$ derived from it. The likelihood that a system of length $x$ is spanned by a continuous path through an interconnected cluster of conductances with arbitrary minimum value, $g$, of the conductance, $W_p(g \mid x)$, is developed from the cluster statistics of percolation theory [38] using the transformation of variables given above and the power-law distribution of local conductances $W(g)$. The cluster statistics are given in terms of $p$ and $p_c$, which can be transformed to $g$ and $g_c$ using Eq. (8) and Eq. (9). The cluster volume, $s$, is transformed [38,67] to a cluster length, $N$, by using the transformation $s \approx N^{1/\sigma v}$, where $v$ and $\sigma$ are universal exponents from percolation theory equal to 0.88 and 0.45, respectively, in a 3D system. Then the product of the cluster statistics and $N^3$ (the Euclidean volume of a cluster of $s \approx N^{1/\sigma v}$ elements) was integrated over $N > x/L$, where $L$ is a particular value of the correlation length (to be discussed below). The reason for the integration is that any cluster of dimension greater than $x$ can span the system. The result for $W_p(g|x)$ was found to be [38-41],

$$W_p(g \mid x) \propto \frac{1}{b} Ei\left[ a \left( \frac{x}{L} \right)^b \right] \tag{10}$$

where

$$Ei[z] = \int_z^\infty \frac{\exp[-y]}{y} dy \tag{11}$$

is the exponential integral, $L$ is a quantity related to the correlation length but influenced by experimental conditions as well [12,40], and the parameters $a$ and $b$ are given by [38-40],

$$a = \left| 1 - \left( \frac{g}{g_c} \right)^{\frac{3-D}{3}} \right|^2 \quad \text{and} \quad b = \frac{2}{v} \tag{12}$$

Here the quantity $v$ defines the divergence of the correlation length in percolation theory [68,69], while the fractal dimension of the pore space $D$ describes the *width* of the power-law local conductance distribution.

At moisture contents less than the porosity, the Young-Laplace equation implies that the conductance distribution is effectively truncated at the largest water-filled pore. Under these circumstances, Eq. (10) for $W_p(g \mid x)$ must be rewritten

$$W_p(g \mid x;\theta) \propto \frac{1}{b} Ei\left[ a\left(\frac{x}{L}\right)^b \left(\frac{\beta - \phi + \theta - \theta_t}{\beta - \theta_t}\right)^b \right] \qquad (13)$$

Here $\theta_t$ is the critical volume fraction (expressed as a moisture content) for percolation while the other parameters retain the same meaning as in Eq. (10). The additional saturation-dependent factor in the argument of the exponential integral is provided by a scaling argument that makes the factor equal to one at saturation and allows for a divergence of the correlation length when $\beta = \phi$ and the moisture content takes on its critical value. Note that $W_p(g \mid x;\theta)$ is defined by Eq. (13) only between the bounds $g_{max}$ and $g_{min}$ corresponding to $r_{max}$ and $r_{min}$. In keeping with the notation that expresses all quantities in terms of the critical conductance, $g_c$, these values are:

$$g_{max} = g_c \left[ \frac{\beta - \phi + \theta}{\beta - \phi + \theta - \theta_t} \right]^{\frac{3}{3-D}} \qquad (14)$$

$$g_{min} = g_c \left[ \frac{\beta - \phi}{\beta - \phi + \theta - \theta_t} \right]^{\frac{3}{3-D}} \qquad (15)$$

Results for the saturated case can easily be obtained by substituting $\theta = \phi$. In the limit of zero moisture content, $g_{max} = g_{min}$, even though Eq. (14) and Eq. (15) really only apply for $\theta \geq \theta_t$. In previous publications [38-40], upper and lower bounds on $g$ in $W_p(g|x)$ were omitted. This is the omission alluded to earlier.

The exponential integral has a logarithmic, not a power-law, divergence at the critical conductance, as given by the asymptotic expansion of the exponential integral function. Thus the power-law form of the local conductance distribution is not relevant to the asymptotic (long-time) behavior of the arrival time distribution, since the pertinent input from this distribution is the logarithmic divergence [38-41]. The more important input to the distribution tail is specified by the divergence of the arrival times at the critical conductance (shown next) and is related to the fractal structure of large clusters as given in percolation theory. Thus, the structure of the medium is of little importance when compared to the relevance of the topology of the optimal (controlling) flow paths as defined through percolation theory [58].

The probability distribution function (pdf) for solute arrival times in a system of length $x$, $W_p(t \mid x;\theta)$ is found using the identity $W_p(t \mid x;\theta) \, dt = g \, W_p(g \mid x) dg$. In order to apply such an identity, one must first have an expression for the typical transit time, $t(g,x)$ of a cluster of length $x$ and controlling conductance, $g$.

The typical time for transport along a one-dimensional path through such a cluster is generated by isolating the effects of cluster topology and the conductance distribution [38]. For the latter, we found the effects on the solute arrival time by using the conductance distribution for the entire medium as a guide. In particular, we terminate that distribution at a minimum value for the path and renormalize it accordingly. For the former we referred to Ref. [59] that related the typical arrival time to the fractal dimensionality of the backbone.

The result of these calculations using the Rieu and Sposito pore-size distribution model for saturated conditions was [38]

$$t(g \mid x) = \left(\frac{x}{L}\right)^{D_b} \frac{Dt_0}{3-D} \frac{1}{(1-\theta_t)^{vD_b-v}} \left[\left(1+\frac{\theta_t}{1-\theta_t}\right)\left(\frac{g_c}{g}\right)^{1-D/3} - 1\right] \left[\frac{1}{\left(\frac{g}{g_c}\right)^{1-D/3} - 1}\right]^{(D_b-1)v}$$

$$\equiv \left(\frac{x}{L}\right)^{D_b} t_g \tag{16}$$

Here the exponent $D_b$ is the fractal dimensionality of the backbone, $\theta_t$ is the critical volume fraction for percolation, $v$ is the exponent describing the divergence of the correlation length, $t_0$ is a typical pore-crossing time, and $x$ is the system size in terms of the length $L$. Since the expression is scaled with $g_c$, which ultimately defines the hydraulic conductivity, $K$, $t_0$ itself should presumably scale with $K^{-1}$, since the characteristic time scale for the fluid flow is inversely proportional to the flow rate. Eq. (16) normalizes the pore-size pdf by applying a condition on the total pore volume, i.e., porosity. For the construction of a 1-D path, however, we now consider it preferable to normalize the relevant pdf in terms of the total number of pores encountered (in Eq. (17), below), explaining the need to use the two different parameters $k$ and $B$ in Eq. (1) and Eq. (4). This we regard as the second minor correction to previous works.

If we now solve for the saturation dependence of $t(g \mid x;\theta)$ for the generalized pore-size distribution model (Eq. 6), the result is,

$$t(g \mid x;\theta) = \left[\frac{x}{L}\right]^{D_b} \frac{Dt_0}{3-D} \left[\frac{\left(\frac{g_c}{g}\right)^{1-D/3}\left(\frac{\beta-\phi+\theta}{\beta-\phi+\theta-\theta_t}\right) - 1}{1-\left(\frac{g}{g_c}\right)^{\frac{D}{3}}\left(\frac{\beta-\phi+\theta-\theta_t}{\beta-\phi+\theta}\right)^{\frac{D}{3-D}}}\right] \left[\frac{1}{(\beta-\phi+\theta-\theta_t)^{(D_b-1)v}}\right] \left[\frac{1}{1-\left(\frac{g}{g_c}\right)^{1-D/3}}\right]^{(D_b-1)v}$$

(17)

In the absence of the factor in the denominator in the first square bracket (from the alternate normalization), Eq. (17) would reduce to Eq. (16) under simultaneous application of the conditions $\theta=\phi$ and $\beta=1$. Note that all the parameters in this equation have a physical meaning. Some, like the exponent for the correlation length, $v$, are believed to be universal for a given flow dimension [68,69]. Others, like the exponent for the fractal dimensionality of the backbone, are known, under a wide range of conditions [58], to have only a limited variability (in particular, with flow dimensionality and whether the percolation problem is random or invasion), but could also have a wider range of values given certain correlation structures for the local conductance distribution [70,71]. The critical volume fraction for percolation will not, in general, be known, but a good estimate can be made from the SWRC, where it can be identified with the minimum water content measured (residual water content) [66]. The fractal dimensionality, $D$, and the parameter $\beta$ can also be extracted from the soil water retention curve.

Application of the probabilistic identity mentioned [38-41] yields an expression for $W_p(t \mid x;\theta)$. The numerical procedure to determine $W_p(t \mid x;\theta)$ is rather complicated, however, since it involves an inversion of Eq. (17) for $t(g \mid x;\theta)$, i.e., solution of $g(t)$. While, for any $x$, $t(g)$ is single-valued, $g(t)$ is not, and either one, two, or even three different limiting conductance values can give the same arrival time [38]. Moreover, the existence of multiple powers of $g$ in $t(g)$ means that inversion requires a numerical solution. As a consequence of the potential multi-valued nature of the solution for $g(t)$, our procedure generates, at small system size, inputs to large values of the arrival time from effects of both small values of the limiting conductance

(e.g., $g \ll g_c$) and tortuous paths (e.g., $g \approx g_c$) near the percolation threshold. Further, for small system sizes, it also allows unusual clusters with limiting conductance value much greater than $g_c$ to provide rapid, relatively straight, paths of solute transport. This leads to a narrowing of the arrival time distribution with increase in system size, $x$, on the short time wing of the distribution.

**Results and Discussion**

In Figure 1, we show the variation in $W_p(t \mid x;\theta)$ which arises from variation in $D$ (essentially a proxy for flow heterogeneity) for a typical 3D porous medium with $\phi = 0.4$, $\theta = 0.4$, $\theta_t = 0.08$ and $\beta = 1$. The larger the fractal dimension is, the more heterogeneous is the medium. In this study we set $x/L$ equal to 50. In Figure 2 we show the effects of variation in $\beta$ for a medium with $D = 2.9$, $\phi = 0.4$, $\theta = 0.1$ and $\theta_t = 0.08$. As can be seen in this figure, the arrival time distribution is rather sensitive to the value $\beta = 0.42$, close to the porosity, corresponding to the case that the pore-size distribution approaches the Tyler and Wheatcraft [63] model. As $\beta$ approaches $\phi$, $r_{min}$ (and also $g_{min}$) tends to zero so that the ratio $g_{max}/g_{min}$ tends to infinity. It was similarly found [66] that the percolation-based hydraulic conductivity model is also sensitive to $\beta$, and values close to porosity change the shape of the hydraulic conductivity function significantly. Since the arrival time distribution is most fundamentally related to the conductance distribution, the impact of $\beta$ on the arrival time distribution, especially for $\beta$ close to the porosity, is not too surprising. In Figure 3 we also show the corresponding variation with $\theta_t$ in a 3-D medium in which $D = 2.9$, $\phi = 0.4$, $\theta = 0.1$ and $\beta = 1$.

For Figures 1 to 3, we set the typical pore-crossing time $t_0$ equal to 1. Generally, as water content decreases, $t_0$ increases inversely proportional to the hydraulic conductivity $(t_0(\theta) \propto 1/K(\theta))$. If one wishes to know the full effects of changes in saturation on an arrival time distribution, this retardation of $t_0$ should be incorporated into the calculation, as we show in Figures 4 and 5. In Figure 4, we show the variation arising from different values of $\theta$ for a typical 2-D porous medium with $D = 1.8$, $\phi = 0.5$, $\theta_t = 0.20$ and $\beta = 1$. As Figure 4 indicates, when the water content decreases from 0.5 (saturation) to 0.3 (close to percolation threshold), the arrival time distribution curve is shifted to the right. We found a time shift of 6 orders of magnitude in a 3-D system in which $D = 2.9$, $\phi = 0.4$, $\theta_t = 0.08$ and $\beta = 1$ (Figure 5) when the saturation was reduced from 1 to 0.25. All of our results for any given choice of percolation type (either random or invasion) and flow dimensionality are consistent with the same slope of the arrival time distribution at large times. However, the peak width does depend on the parameters investigated. This means that typical experimental investigations, which cannot realistically generate an arrival time distribution over 5-10 orders of magnitude of time, may not be able to return the universal slopes that we obtain. Further, experiments typically report breakthrough curves from step function solute pulses that relate to the integral of $W_p(t \mid x;\theta)$. For this reason, we investigated the variability of the slope that one might extract from experimental results for a breakthrough curve.

The results for the slope in a 2D system with $\phi = 0.6$ and different $D$, $\theta_t$ and $\beta$ values are presented in Table 1. We found a range of -2.020 to -1.444 and -4.040 to -3.362 for the random and invasion percolation classes, respectively, which includes the value of -1.58 found [42] in 2D simulations and already generated by our theory in [38]. Furthermore, the smallest values for the slope that we obtained were appropriate for the narrowest pore size distributions, generally compatible with simulation [42] (which incorporated only topological disorder and no pore-size disorder). The minimum and maximum slopes calculated for a 3D medium with $\phi = 0.4$ and different $D$, $\theta_t$ and $\beta$ values are summarized in Table 2 for two random and invasion percolation classes. A range of -2.118 to -1.453 and -3.218 to -2.385 was found for random and invasion percolation classes, respectively. The results show that our model is also able to generate a Gaussian distribution, which occurs when the slope of the distribution tail is steeper than -3, as discussed in Ref. [13]. Under those conditions the existence of the first two moments in time allows application of the central limit theorem.

Another important result to show is the evolution of $W_p(t \mid x;\theta)$ with increasing transport distance, or system size. Figure (6) shows that as the system size increases, the time required for particles to move through the system increases rapidly as well, and the arrival time distribution is shifted to the right.

The spatial distribution at an instant in time, $W_p(x \mid t;\theta)$, is obtained similarly to the procedure for generating $W_p(t \mid x;\theta)$, as described in [38]. $W_p(x \mid t;\theta)$ can be used to obtain the variance of the solute distribution as a function of time, Var($t$), as well as derived quantities such as the longitudinal dispersion coefficient, $D_l(t)$, and the dispersivity, $\alpha_l(x)$. The former is defined as one half the time derivative of Var($t$), while the latter is given as $D_l(t)$ /<$u$>, where <$u$> is the mean solute velocity. If Var($t$) is a power of the time, then the time derivative is proportional to Var($t$)/$t$. It is usual, but not necessary, to represent the longitudinal dispersion coefficient as a function of time. The definition of the variance in terms of the difference between <$x^2$> and <$x$>$^2$ makes it convenient to represent the dispersivity very simply as [12],

$$\alpha_l(x) = \frac{\langle x^2 \rangle - \langle x \rangle^2}{\langle x \rangle} \tag{18}$$

In this procedure, it is typical, though not necessary, to represent the dispersivity as a function of *x*, as written. Experimental results are also typically given in this form.

Conversely, we can use $W_p(t \mid x; \theta)$ to calculate *Var(x)*. The result that *Var(x)* is proportional to $\langle x \rangle^2$, as established in dispersive transport in semiconductors [13,14], leads immediately to a dispersivity that is proportional to $\langle x \rangle$, as is also known from groundwater studies [72]. While it has been reported [73,74] that the dispersion coefficient is typically a small power (sublinear) of time and that the dispersivity is roughly a linear function of *x*, the necessary consequence (as shown in Figure 7), is that the time for particles to reach a given distance *x* is a superlinear function of *x*, has been reported only once [75] (to our knowledge) for solute transport in porous media, but is well-known in dispersive transport [45-50]. In particular, we have then *Var(x)* ∝ $\langle x \rangle^2$ and *Var(x)* ∝ $t^{1+\delta}$, with $\delta \ll 1$ giving immediately $t \propto x^{2/(1+\delta)}$. When $\delta \to 0$, $t \propto x^2$, for example.

Figure (7) shows the length dependence of typical system crossing times as derived from numerically inverting $\langle x(t) \rangle$. The exponents are largely determined by the assumed scaling of time with the fractal dimensionality of the backbone (and the variation of this scaling with the dimensionality of transport has been confirmed in [12]) but a variability of roughly 10% is introduced by finite-size effects such as the narrowing of the distribution with increasing transport distance. It has already been shown [41, under review] that the results from the simpler fractal model (e.g., [62]) describe perfectly the observed variability of this scaling exponent in dispersive transport.

The results for the length dependence of the dispersivity show some significant variability, but this variability nearly disappears in the limit of large transport distance (Figure (8)). We argue that the regime where the various predictions coincide exhibits universal characteristics. At smaller length scales, variation in the width of the conduction distribution or dimensionality of flow paths, or class of percolation problem, produces a rather wide range of predicted values of the dispersivity, in agreement with experiments from Pachepsky et al. [17] and Refs. [76-83]. Since the excellent agreement has already been discussed in detail elsewhere [12,40], we will not go into detail here. Nevertheless, we should emphasize that the introduction of the limits on the controlling conductance ($g_{min}$ and $g_{max}$) from the truncated conductance distributions has restricted the range of predicted values (compared with Refs. [12, 40]) at large values of the dispersivity and small length scales. However, it may have been unrealistic to expect a single model (a monomodal power-law conductance distribution) to reproduce all the observed data. Further, it is important to reemphasize that the variability as predicted was generated by choosing a single length scale, $L$, of 1 meter. The only possibility that we could envision to make this single length scale relevant to all experiments (except the micromodel experiments at scales of millimeters) was that it was largely determined by the experimenter, probably in the choice of the initial volume of solution [12,40].

**Comparison with experiments**

We turn to the dependence of the breakthrough curve on saturation. Since the experimental arrival time distribution is obtained from the derivative of the experimental breakthrough curve, it proved necessary to develop a method to minimize the impacts of discretization uncertainty. The quantity $C/C_0$ (in which $C$ is the solute concentration and $C_0$ is the initial concentration) is not directly the distribution of arrival times, but rather, related to $\int W(t)dt$. Thus, we first invert the measured time-based breakthrough curve to 1-$C/C_0$ vs. time, fit it to an appropriate function (e.g., a series of Gaussian functions) then differentiate the corresponding function and multiply that by -1 to generate the arrival time distribution $W(t)$.

We address here primarily additional experimental comparisons made possible by the new theoretical development, but also show that these changes do not affect importantly the most striking results obtained previously.

*Prediction of arrival time distribution at different saturations from the breakthrough curve measured at saturation*

Three soil experiments measured by Jardine et al. [84] were used to evaluate how the predicted arrival time distribution compares with the experimental measurements. The physical properties of each experiment are presented in Table 3. The interested reader is referred to the original article published by Jardine et al. [84] for more information on the experiments. We should point out that the soil water retention curve is not available for these datasets.

Our first task is to discuss the values of the percolation exponents that should be used to make our predictions of the arrival time distribution. Since the experiments were performed in three dimensions, we use values for the correlation length and the fractal dimensionality of the backbone for three dimensions. The case in which the medium is fully saturated clearly requires that we choose $D_b = 1.87$ from random percolation, since no remnants of the topology of the water invasion remain. In order to use the value of $D_b = 1.46$, the appropriate percolation problem must be bond, trapping, invasion percolation [58]. The experiments were performed under conditions of drying. Sahimi [57] argues that drying is a bond invasion percolation problem and that the incompressibility of water leads to trapping. One might thus assume that $D_b = 1.46$ would be the appropriate choice for all saturations less than one for which a connected path of water-filled pores exists. However, such a hypothesis does not quite work as is seen by our comparisons with experiment.

By fitting our model to the numerically calculated arrival time distribution obtained from experiments (as explained above) for the saturated case ($h = 0\ cm$) in which $D_b = 1.87$ and $\theta = 0.549$ (Figure 9), we found values of $D = 2.966$, $\theta_t = 0.15$ and $\beta = 0.8$ as the best fit. In fact, $D_b = 1.87$ describes the saturated case very well. Then, we used the same $D$, $\theta_t$ and $\beta$ values to predict the arrival time distribution for unsaturated cases ($h = 10$ and $15\ cm$). The results shown in Figure (9) indicate that $D_b = 1.46$ (from invasion percolation) is an excellent choice for tension $h$

= 15 *cm* (at saturation 93%). But the intermediate case at 10 *cm* (saturation 97%) could not be described well by the value of the fractal dimensionality of the backbone from invasion percolation. We found that the $D_b$ value from random percolation was more appropriate than that of invasion percolation in prediction of the arrival time distribution (Figure 9). Although the experimental arrival time distribution curve actually appears to conform to the random percolation prediction for intermediate time scales, the prediction was not accurate at large time scales. Thus, at saturation 97%, the experimental results still more closely resemble random percolation and it is not until the saturation drops to 93% that invasion percolation clearly becomes a superior choice. The results obtained indicate that the air entry value $h_{min}$ might be in the range of 10 to 15 *cm*.

We also found that the typical pore-crossing time $t_0$ should inversely scale with hydraulic conductivity $t_0 \propto K^{-1}$ in which $K$ is [66]:

$$K(\theta) = \begin{cases} K_s \left[ \dfrac{\beta - \phi + \theta - \theta_t}{\beta - \theta_t} \right]^{D/(3-D)} & \theta_d \leq \theta < \phi \\ K_s \left[ \dfrac{\beta - \phi + \theta - \theta_t}{\beta - \theta_t} \right]^{D/(3-D)} \left( \dfrac{\theta - \theta_t}{\theta_d - \theta_t} \right)^2 & \theta_t \leq \theta \leq \theta_d \end{cases} \quad (19)$$

where $K_s$ is the saturated hydraulic conductivity, and $\theta_d$ is the cross-over point on the hydraulic conductivity curve which recognizes percolation scaling from fractal scaling [66]. Note that the second form of Eq. (19), in the case that $\beta = \phi$, generates the known form of non-universal scaling of the hydraulic conductivity derived by Balberg [85]. As can be seen in Figure 9, the time value at the peak of arrival time distribution ($t_p$) for $h$= 0, 10, and 15 *cm* is about 20, 200, and 3000 *min*, respectively. Therefore, the ratios $t_p(h = 10)/t_p(h = 0)$, $t_p(h = 15)/t_p(h = 10)$, and $t_p(h = 15)/t_p(h = 0)$ would be 10, 15, and 150, respectively. The ratios $K(\theta = 0.549)/K(\theta = 0.533)$ = 8.8, $K(\theta = 0.533)/K(\theta = 0.513)$ = 16.4, and $K(\theta = 0.549)/K(\theta = 0.513)$ = 144.1 were calculated using Eq. (19) with $D$ = 2.966, $\theta_t$ = 0.15 and $\beta$ = 0.8. Thus, the same set of parameters yields both the appropriate shapes of the arrival time distribution and the scaling of the most likely arrival time with saturation.

*Prediction of arrival time distribution at different saturations from measured soil water retention data*

We used 6 Hanford sediment experiments measured by Cherrey et al. [86] at different saturations e.g., $\theta$ = 0.4 (saturation), 0.237, 0.204, 0.172, 0.139 and 0.126. Table 4 shows the selected properties of each experiment, and more information can be found in the paper by Cherrey et al. [86]. The measured soil water retention curve shown in Figure 10 indicates that the critical water content for percolation $\theta_t$ is about 0.074. Fitting the soil water retention model developed in this study Eq. (6) to the measured data yielded $D$ = 1.95, $\beta$ = 0.4 and $h_{min}$= 4.75 *cm* (Figure 10).

Although we predict the dependence of the arrival time distribution on saturation very well for the unsaturated medium (Figure 12) in this series of experiments, the predicted arrival time distribution for complete saturation (Figure 11), which was narrower than at unsaturated conditions, did not match our predictions at all when we used the exponents from 3D random percolation. The authors [86], however, mentioned that complications due to "wall flow," a relatively common problem in Hanford sediments, could be excluded. Given that the particle size distribution was rather coarse, however, which is known to be a contributing cause for wall flow, we decided to investigate the possibility that the dispersion experiment performed under ostensibly saturated conditions could be influenced by this phenomenon.

The problem in wall flow is that the boundary of the core wall and the medium may be distorted by the presence of many coarse particles to produce a region of higher porosity and thus preferential flow. Owing to the large pores, however, this portion of the medium is difficult to maintain at saturation. Since it is such a small fraction of the medium, even if it is only 80% saturated, the medium generally may be at 99% saturation, which is exceedingly difficult to distinguish from 100% saturation. Since the boundary might thus be unsaturated and its configuration is two-dimensional, we tried using invasion percolation exponents in two dimensions. This produced a much closer correspondence with experiment, except at large times, where we still overestimate particle arrivals.

In the unsaturated cases, we also somewhat overestimate the arrival time distribution at long time scales. There are two possibilities: (1) At long times diffusion might eliminate the relevance of highly tortuous paths to dispersion as Hunt and Skinner [39] pointed out (Figure 6 of the Hunt and Skinner [39] article), (2) There could be a relevance of a bimodal distribution of local conductances. Since the Peclet number values (presented in Table 2 of Cherrey et al. [86]) are in the range of 5 to 300, convection dominates dispersion, but the effect of diffusion may be important [57]. Furthermore, recently Bijeljic et al. [87] generalized the continuous time random walk approach for media showing two distinct regimes and demonstrated its relevance. For the saturated case, we might appeal to the distinct portions of the medium possibly implicated above. For the unsaturated experiments, we revisit the soil water retention curve. In particular, reanalyzing the soil water retention curve plotted on log-log scale indicated two fractal regimes

($D_1$ = 2.030 and $\beta_1$ = 0.49 for the first regime $0.12 \leq \theta < 0.4$, and $D_2$ = 2.943 and $\beta_1$ = 0.53 for the

second regime $0.07 < \theta < 0.12$). Trying to verify/falsify either of these hypotheses is beyond the scope of the present study, however.

Note that all the parameters except the fractal dimensionality of the backbone and the exponent for the correlation length, whose values were stipulated from percolation theory, were derived from fitting the water retention curve. Thus, we had no free parameters to adjust. Furthermore, we wish to emphasize that, for the unsaturated cases, the predicted scaling of the peak arrival times with saturation was in each case within 50% of the observed scaling, and in two of the four cases within 10%. As we show in Figure 12, the time value at the peak of arrival time distribution ($t_p$) for $\theta$ = 0.237, 0.204, 0.172, 0.139, and 0.126 is near 1.3, 2.8, 9.3, 66.5, and 300 $min$, respectively, meaning that the ratios $t_p(\theta = 0.204)/t_p(\theta = 0.237)$, $t_p(\theta = 0.172)/t_p(\theta = 0.204)$, $t_p(\theta = 0.139)/t_p(\theta = 0.172)$, and $t_p(\theta = 0.126)/t_p(\theta = 0.139)$ are 2.15, 3.32, 7.15, and 4.51, respectively. The ratios $K(\theta = 0.237)/K(\theta = 0.204) = 2.39$, $K(\theta = 0.204)/K(\theta = 0.172) = 2.97$, $K(\theta = 0.172)/K(\theta = 0.139) = 4.87$, and $K(\theta = 0.139)/K(\theta = 0.126) = 2.36$ were calculated using Eq. (19) with $D$ = 1.95, $\theta_t$ = 0.074 and $\beta$ = 0.4. Thus, in this case, all facets of the experimental results at saturations less than 1 were reasonably well predicted from theory, while the saturated case may have provided the opportunity to diagnose an experimental complication.

We also note that the predicted peak shape was dominated by universal exponents from percolation theory, even though the dependence of the peak arrival time was obtained from non-universal scaling of the hydraulic conductivity (as a consequence of the coincidence that $\beta=\phi$). This result is in contrast to a recent publication of Sahimi [71], in which the non-universality of the hydraulic conductivity requires a related non-universality in the distribution of arrival times. While analysis of this particular experiment appears to favor our treatment, we regard this issue as unresolved.

**Summary**

We have applied a new method for calculating dispersion properties of solutes in flow through porous media to generate explicit dependence on saturation and porosity. The calculation was originally tailored [38] to media with wide ranges of local conductances, for which critical path analysis is the appropriate percolation-theoretical framework for flow and transport. In the current adaptation to generate the saturation dependence of the dispersion, we had to make the calculations suitable for structural percolation controls, i.e., direct topological constraints. The theoretical development is strongly influenced by work of the Eugene Stanley group [59], which shows that characteristic system crossing times scale with system length to the fractal dimensionality of the backbone, $D_b$. We did not use the scaling function for arrival times used in [59] and related publications, however. That distribution was proposed for use in a binary medium, in which the constituents are either highly, or weakly, permeable. In our case we have a continuous range of local permeabilities. The principle variability in our distribution of arrival times arises from the variability in clusters as described by the value of the controlling conductance, rather than in the variability of the paths across a given cluster. Consequently, it may be somewhat surprising that our method generated [38] such close correspondence to simulations performed [42] at the percolation threshold and without pore-size variations. This success was part of the motivation for extending the existing theoretical results to account explicitly for the effects of variable saturation, allowing tuning of a system over a range that extends far from the percolation threshold all the way to critical percolation.

Our method had already been shown to predict the variability of the dispersivity as a function of length scale over 10 orders of magnitude of length scale [12,40], as well as to predict the length dependence of the system crossing times for non-equilibrium electronic transport (dispersive transport) in disordered semiconductors and polymers [12]. We wished to determine whether our theory generated significant variability in the exponent describing the (approximate) power-law tail of the distribution. Here, we have shown that considerable variability in such an experimentally extracted slope could be expected, and that this variability would arise from the weaker effects of the particular heterogeneity found in given experimental systems. We have also shown here that in one soil [84], at least, the variation in the tail of this distribution as a function of saturation is compatible with the variation in $D_b$ associated with a cross-over from the relevance of invasion percolation at incomplete saturation (93%) to random percolation at complete saturation (100%). In particular, with only the expected change in $D_b$ and (more minor) changes from the explicit saturation dependence, we were able to generate both experimental curves, holding all other parameters constant. It should be mentioned, however, that at 97% saturation, the agreement with random percolation was not perfect at all time scales. We investigated a second system (a Hanford site sediment [85]) and found that under unsaturated conditions the dependence of the arrival time distribution, including the time scale, was well predicted over a saturation range of 16% to 50%. Note that in this case, all parameters (except the fractal dimensionality of the backbone and the exponent for the correlation length) were obtained from comparison with soil water retention data.

Determining the explicit saturation dependence has the added advantage of allowing an evaluation of the relative importance of the spatial variability of flow velocities compared with explicit structural constraints from percolation theory. Our results show very little difference between the effects of flow and topological heterogeneity, as foreseen by Sahimi [56,57]. However, they do show the large impact of incomplete saturation, representable in terms of a contrast between the applicability of random vs. invasion percolation [57,58], and this contrast is apparently verified here by comparison with experiments.

The strong dependence of the arrival time distribution width on the exponents of percolation theory may also allow the diagnosis of experimental complications, such as wall flow. Such preferential flow, by changing the flow dimensionality from three dimensions to two, and by making the relevant portion of the medium unsaturated, can change the relevant fractal dimensionality of the backbone in percolation theory from 1.87 to 1.21. Such a large diminution in the tortuosity of solute travel paths would show up as a large reduction in the width of the peak of the arrival time distribution. Consequently, in contrast to media without wall flow, where the opposite tendency is observed, the saturated medium will have a narrower distribution of arrival times than the unsaturated medium. Our analysis indicates at least the possibility that wall flow was present in the Hanford sediments that we investigated.

*Acknowledgement*

This research is supported by the US Department of Energy (DOE) Biological and Environmental Research (BER) through Subsurface Biogeochemical research (SBR) Science Focus Area (SFA) program at Pacific Northwest National laboratory (PNNL). We acknowledge the Pacific Northwest National Lab (PNNL) financial support from Battelle contract 154808. We are thankful to Philip M. Jardine and Markus Flury for providing the measured data used in this study. We are also grateful for interactions with John Zachara, Chongxuan Liu, Robert P. Ewing, and an anonymous reviewer.

Figure Captions

Figure 1

Dependence of the arrival time distribution on fractal dimensionality, $D$, of the pore space in the generalized pore-size distribution model for a typical 3D porous medium with $\phi=0.40$, $\theta=0.40$, $\theta_t=0.08$, $x/L=50$, $t_0=1$, $\beta=1.0$ and $D_b=1.87$ (random percolation). The width of the peak increases with increasing $D$, which represents increasing heterogeneity in flow, but the slope at long times remains the same. However, if the slope is estimated in the vicinity of the peak, it will appear to diminish slightly with increasing heterogeneity.

Figure 2

Dependence of the arrival time distribution on the parameter $\beta$ in the generalized pore-size distribution model for a typical 3D porous medium with $D=2.90$, $\phi=0.40$, $\theta=0.10$, $\theta_t=0.08$, $t_0=1$, $x/L=50$ and $D_b=1.87$ (random percolation). As $\beta$ is reduced from 1 towards the porosity (accompanying a transition from the Rieu and Sposito model to the Tyler and Wheatcraft model), the peak width narrows.

Figure 3

Dependence of the arrival time distribution on the critical moisture content for percolation, $\theta_t$, in the generalized pore-size distribution model for a typical 3D porous medium with $D=2.90$, $\phi=0.40$, $\theta=0.10$, $x/L=50$, $t_0=1$, $\beta=1.0$ and $D_b=1.87$ (random percolation). With increasing $\theta_t$ a slight narrowing of the peak occurs.

Figure 4

Dependence of the arrival time distribution on saturation in the generalized pore-size distribution model for a typical 2D porous medium with $D=1.80$, $\phi=0.50$, $\theta_t=0.20$, $x/L=50$, $\beta=1.0$ and $D_b=1.87$ (random percolation). A slight narrowing occurs with increasing saturation.

Figure 5

Dependence of the arrival time distribution on saturation in the generalized pore-size distribution model for a typical 3D porous medium with $D=2.90$, $\phi=0.40$, $\theta_t=0.08$, $x/L=50$, $\beta=1.0$ and $D_b=1.87$ (random percolation). A slight narrowing occurs with increasing saturation.

Figure 6

Dependence of the arrival time distribution on length in the generalized pore-size distribution model for a typical 3D porous medium with $D=2.90$, $\phi=0.40$, $\theta=0.30$, $\theta_t=0.08$, $\beta=1.0$ and $D_b=1.87$ (random percolation).

Figure 7

Dependence of the typical system crossing time on average length for a typical 3D porous medium with $\phi=0.40$, $\theta=0.10$, $\theta_t=0.08$, $\beta=1.0$ and $D_b=1.87$ (random percolation). The range of fractal dimensionalities from $D=1$ to $D=2.5$ produced insufficient change in the predicted scaling to show up clearly in this figure.

Figure 8

Comparison of dispersivity values of over 2200 experiments from Pachepsky et al. [17] and Refs. [76-83] with predicted dependences for reasonable variations in model parameters. Note that some of these experimental values are the same ones reported in Ref. [12]; it was important to check here whether the performance of the theory was degraded by the correction in the limits of $W(g|x)$, as well as to investigate any resulting changes from model variability (PSF vs. RS). The typical values of $\phi=0.60$, $\theta=0.60$, $\theta_t=0.30$, $\beta=1$, and $\phi=0.3$, $\theta=0.3$, $\theta_t=0.08$, $\beta=1$ were used in 2-D and 3-D systems, respectively.

Figure 9

Comparison of the fitted and predicted arrival time distribution model to the Jardine et al. [84] experiments with a choice of the backbone fractal dimensionality from random percolation consistent with complete and close to saturation ($h=0$ and 10 *cm*) and invasion percolation consistent with incomplete saturation ($h=15$ *cm*) resulting from entrapped air. $D=2.966$, $\theta_t=0.15$ and $\beta=0.8$ were found by fitting the model to the saturated arrival time distribution.

Figure 10

The measured soil water retention curve from Ref. [86] and fitted retention model Eq. (6) to the measured data with $D=1.95$, $h_{min}=4.75$ *cm*, $\beta=0.4$ and $R^2=0.96$.

Figure 11

Comparison of the prediction for the arrival time distribution with experiment from Cherrey et al. [86] at saturation for parameters consistent with the soil water retention data ($D=1.95$, $\phi=0.4$, $\theta_t=0.075$ and $\beta=0.4$). The exponents from 3-D random and 2-D invasion percolations are $v=0.88$ and $D_b=1.87$ and $v=4/3$ and $D_b=1.46$, respectively.

Figure 12

Comparison of the prediction for the arrival time distribution with experiment from Cherrey et al. [86] at different water contents $\theta=0.237$, 0.204, 0.172, 0.139 and 0.126 for parameters consistent with the soil water retention data ($D=1.95$, $\phi=0.4$, $\theta_t=0.075$ and $\beta=0.4$) and a choice of the backbone fractal dimensionality from invasion percolation $D_b=1.46$.

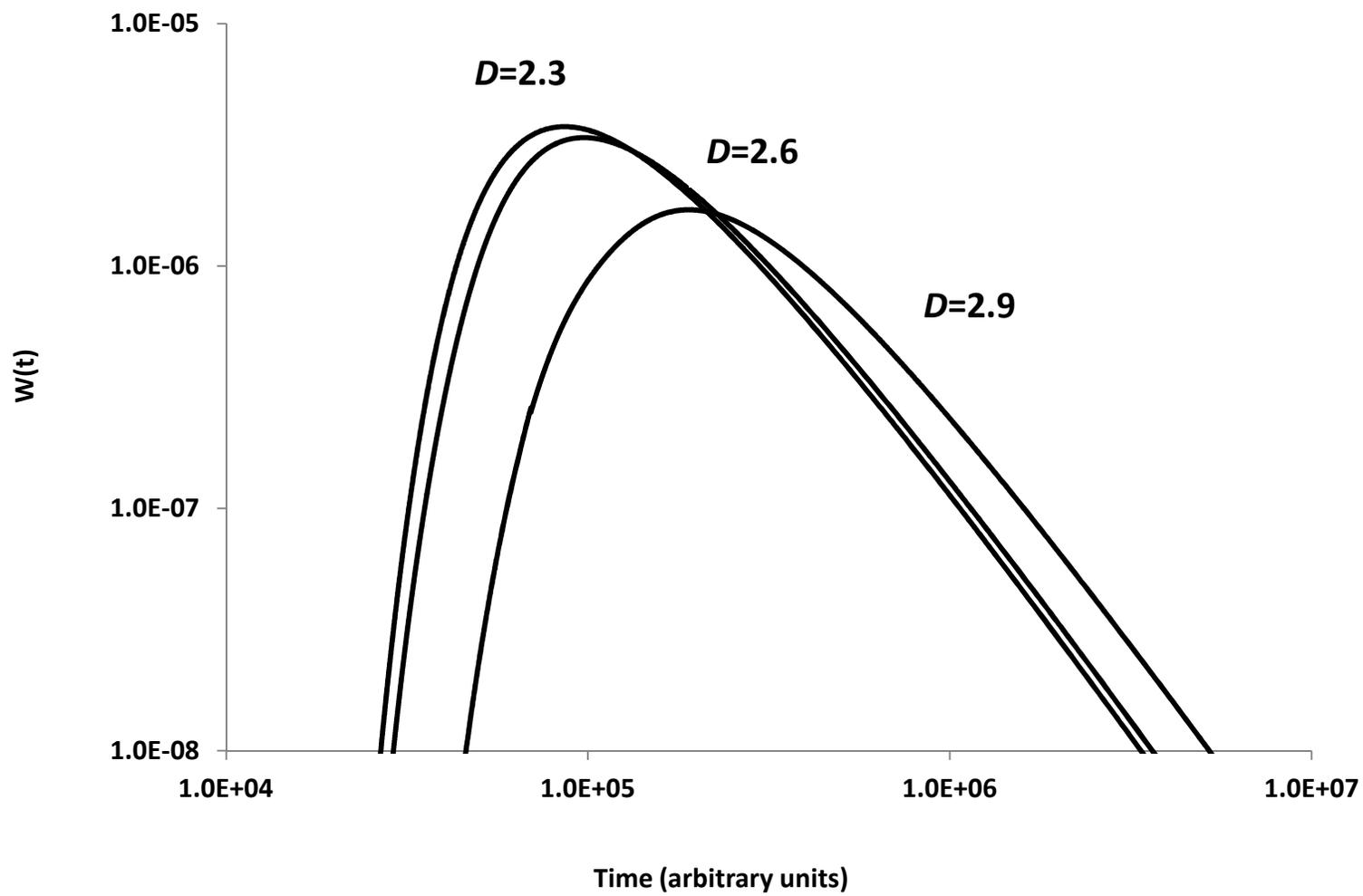

Figure 1

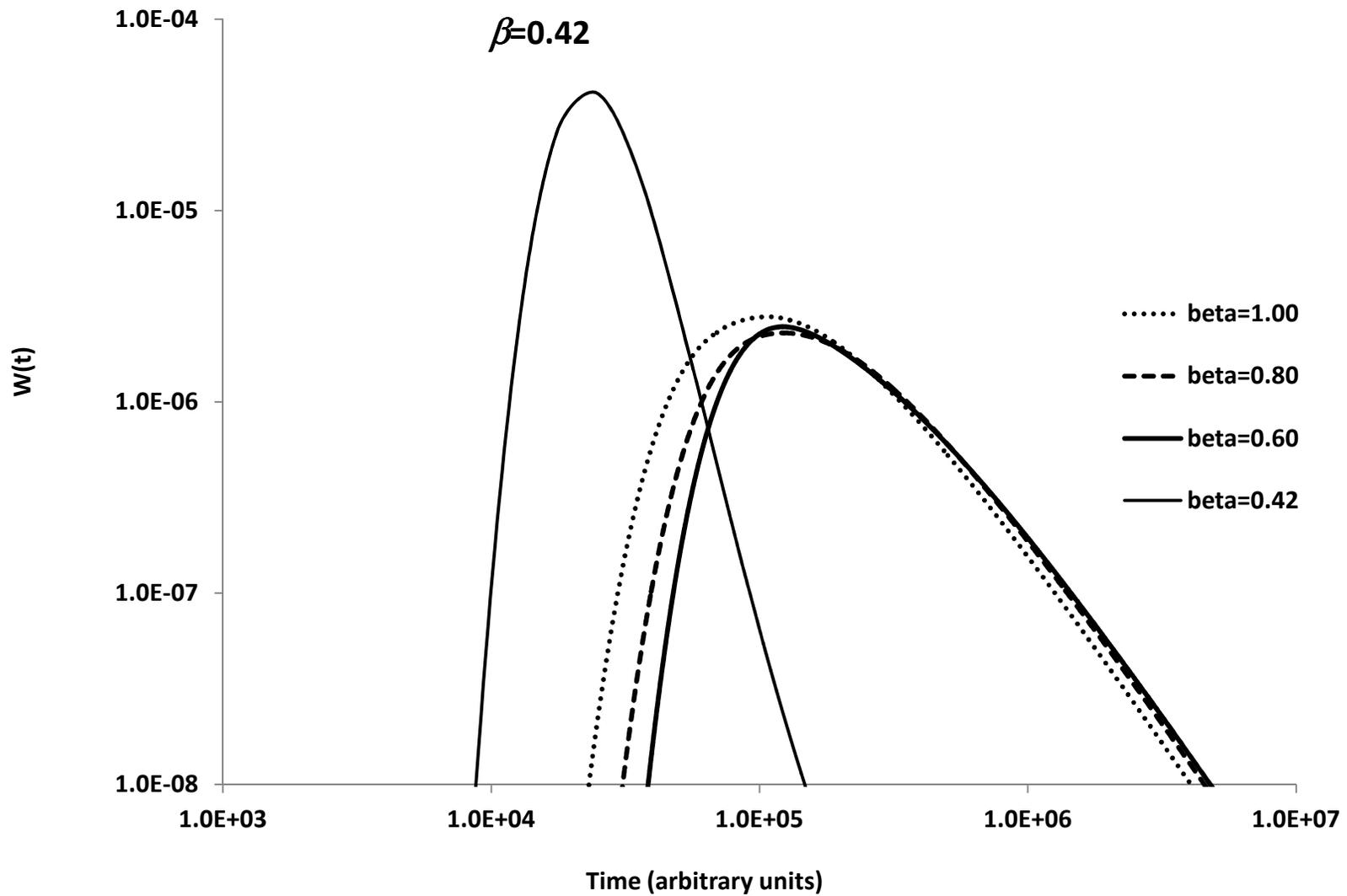

Figure 2

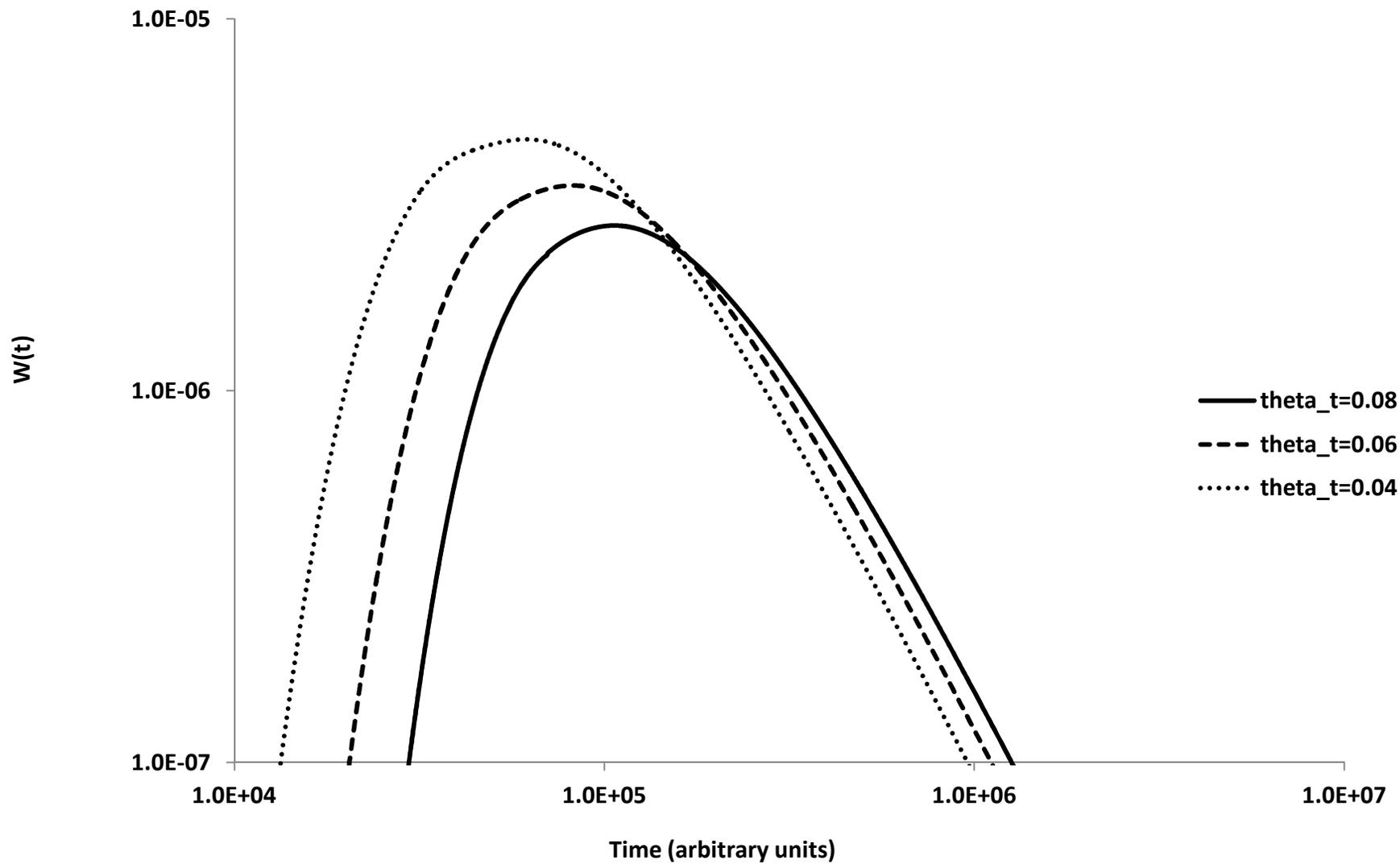

Figure 3

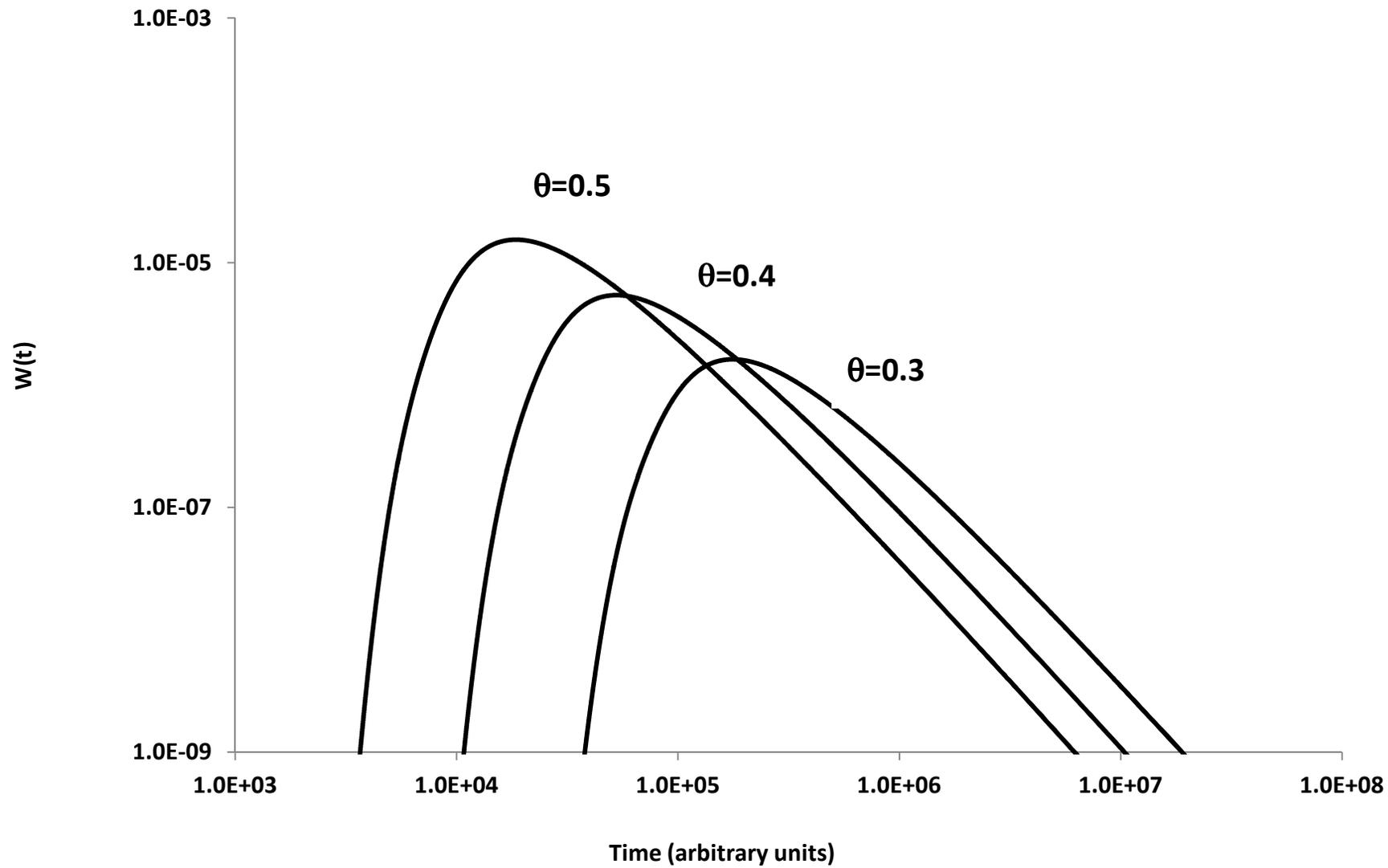

Figure 4

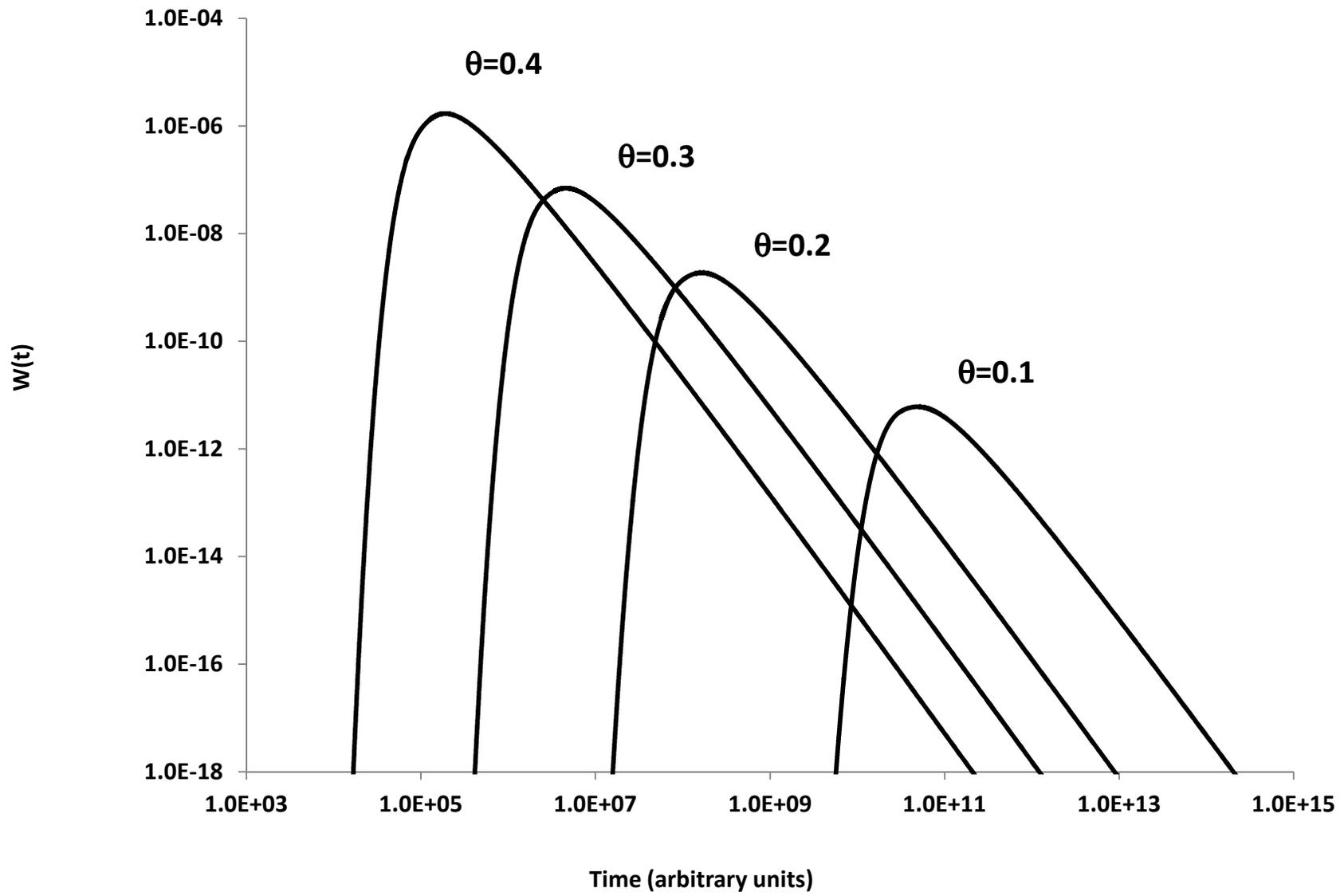

Figure 5

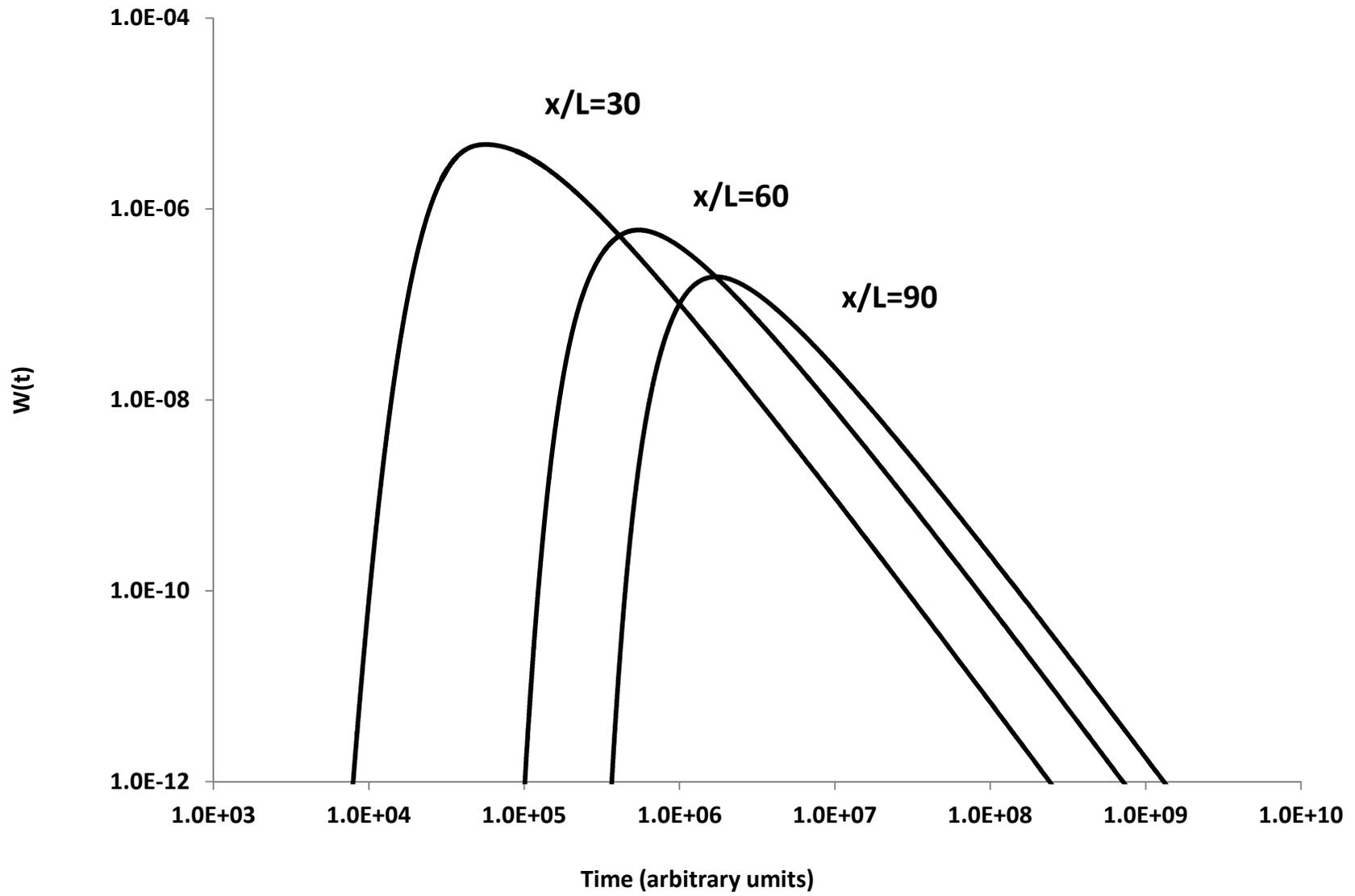

Figure 6

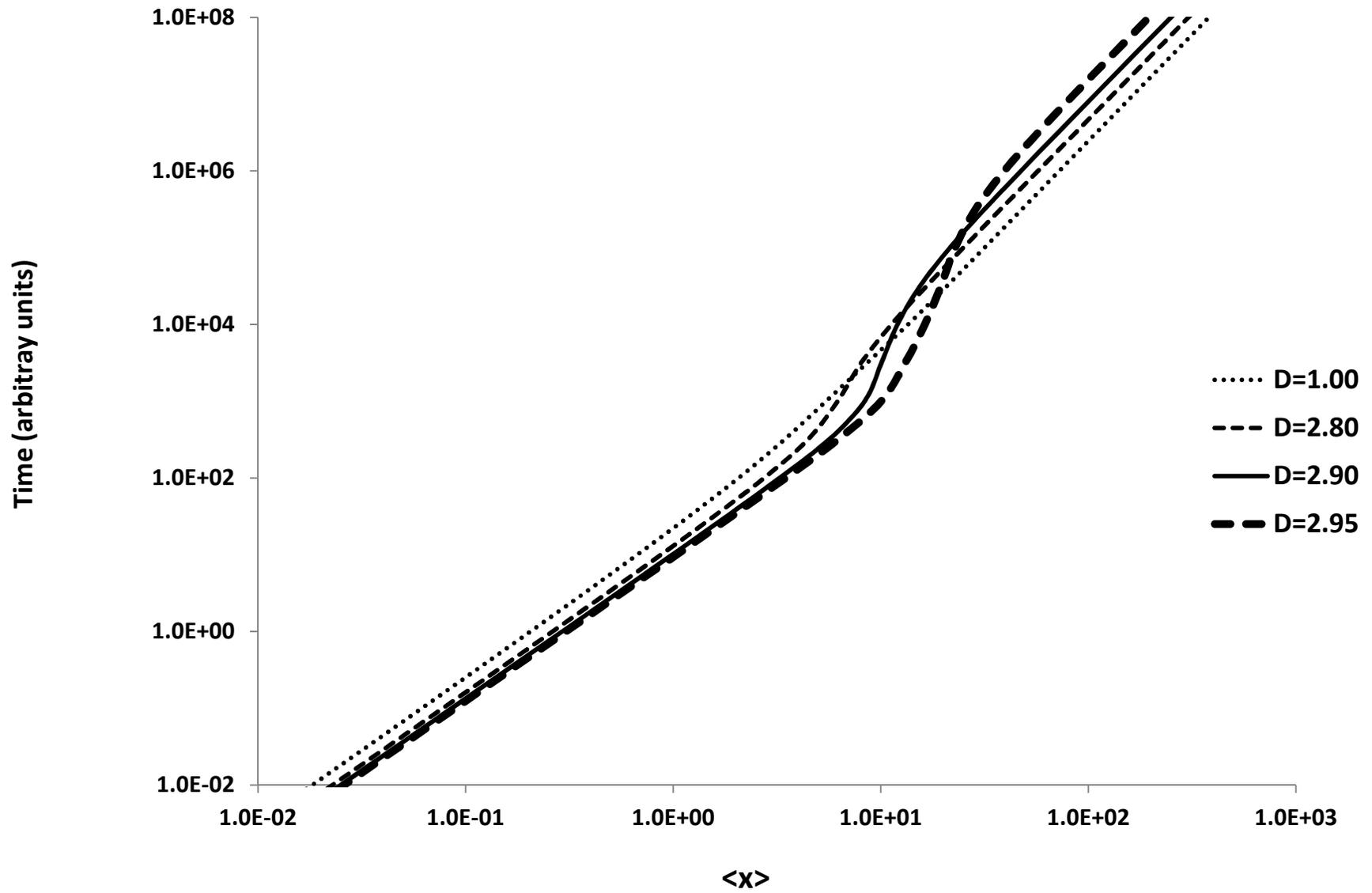

Figure 7

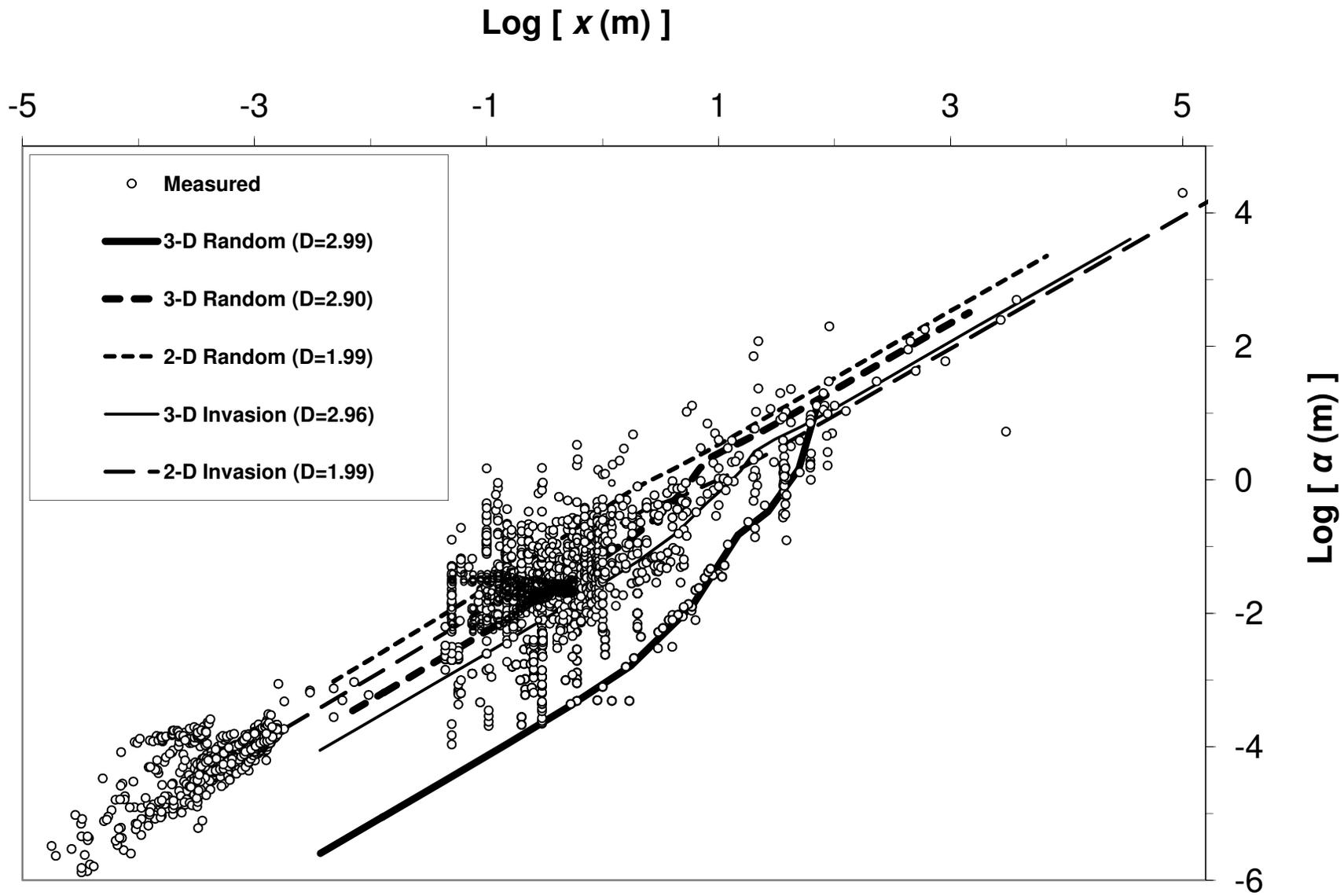

Figure 8

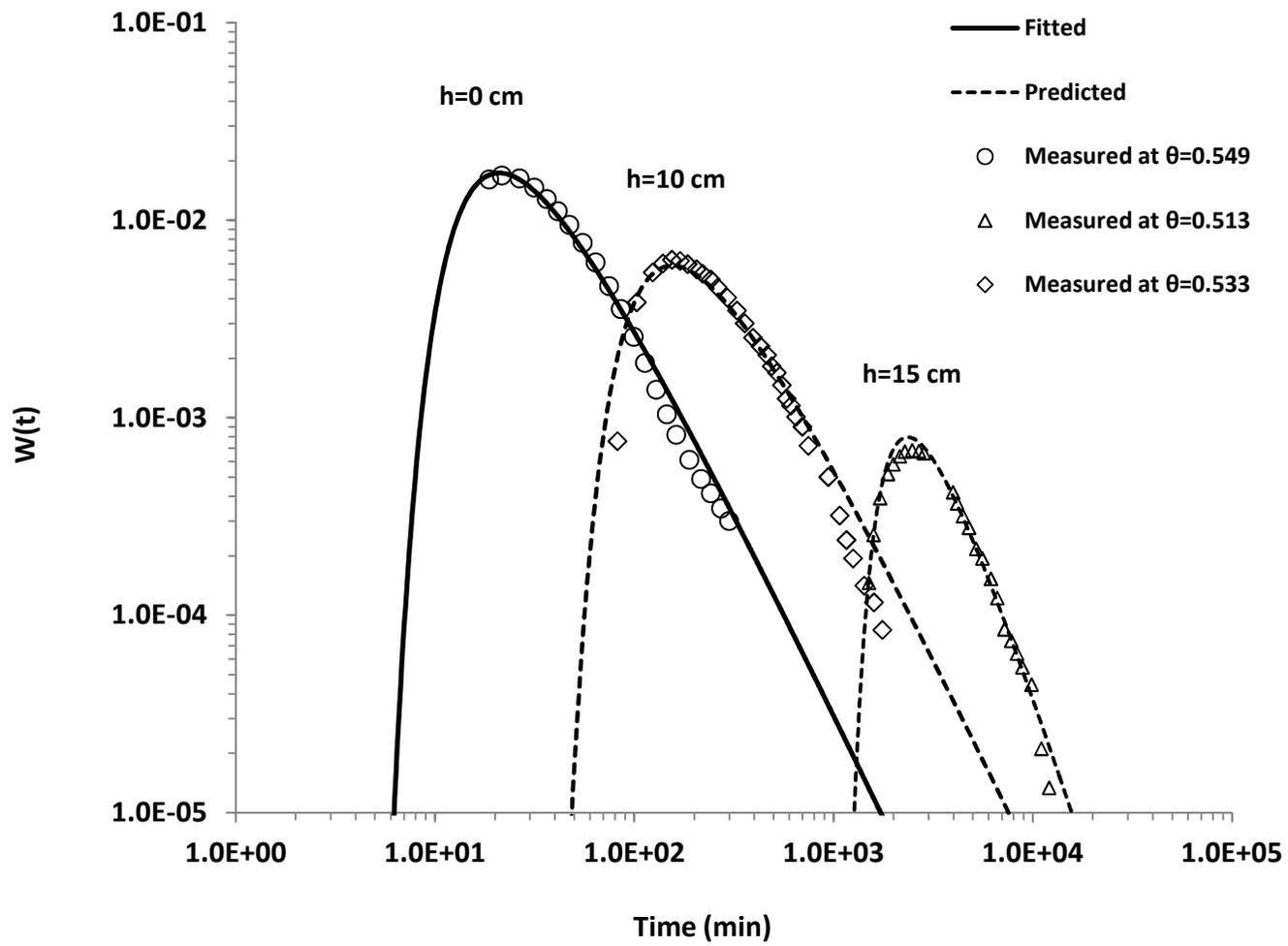

Figure 9

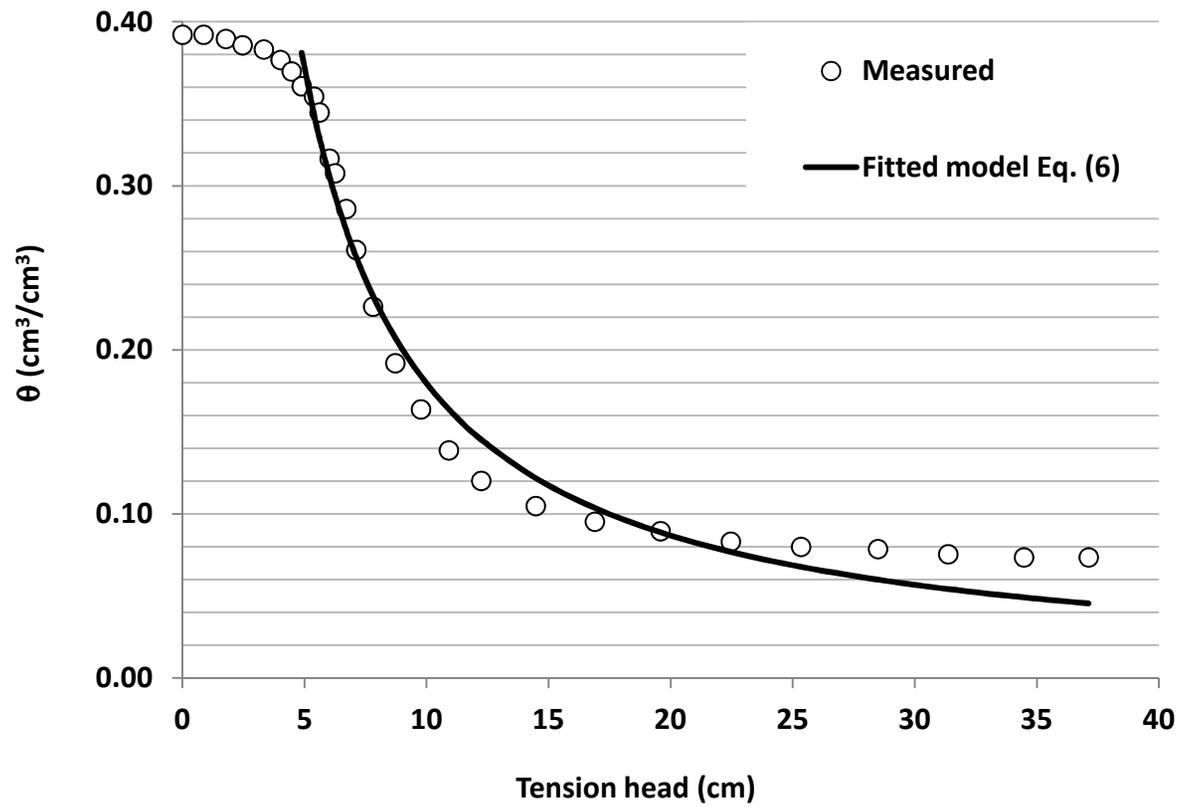

Figure 10

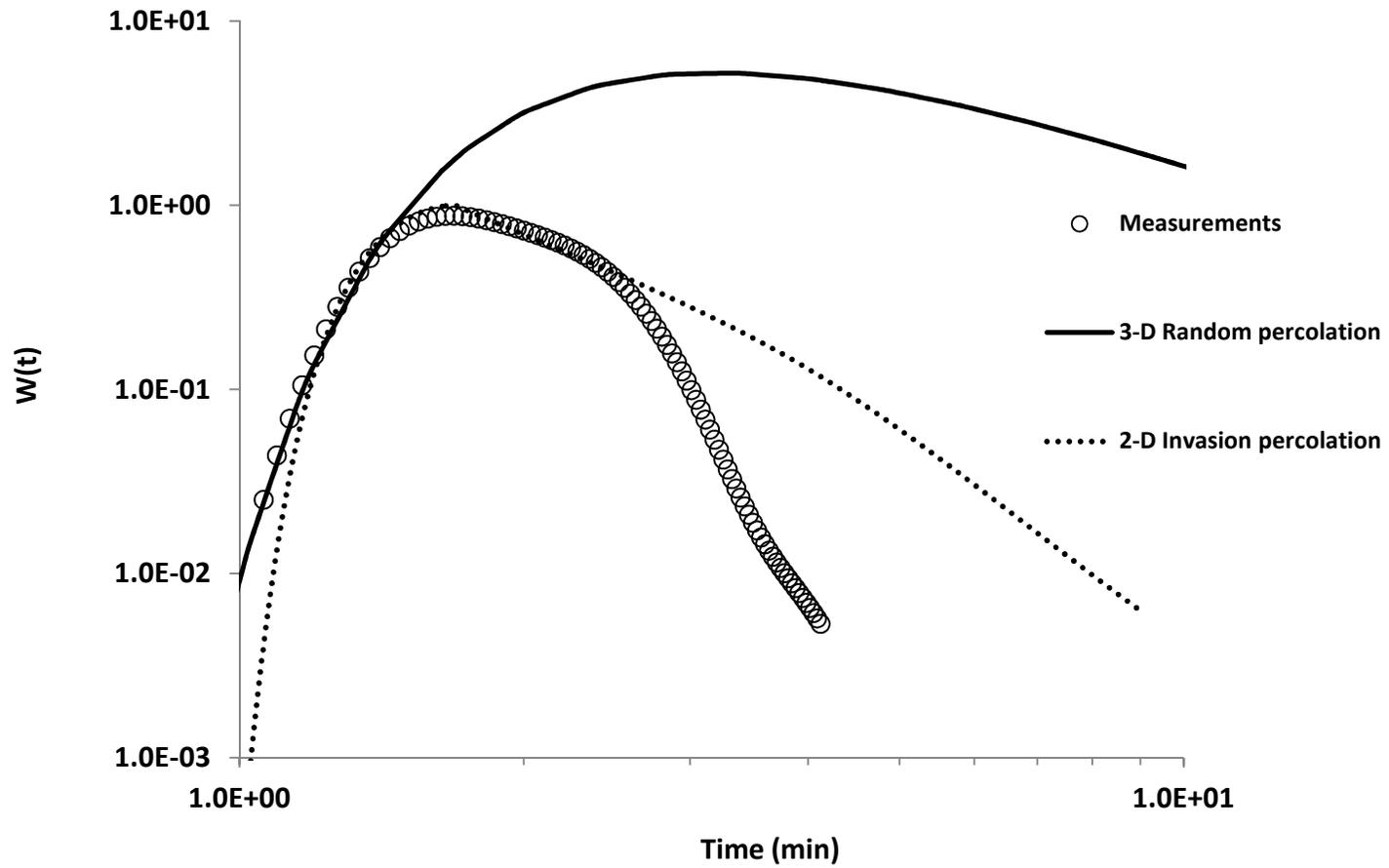

Figure 11

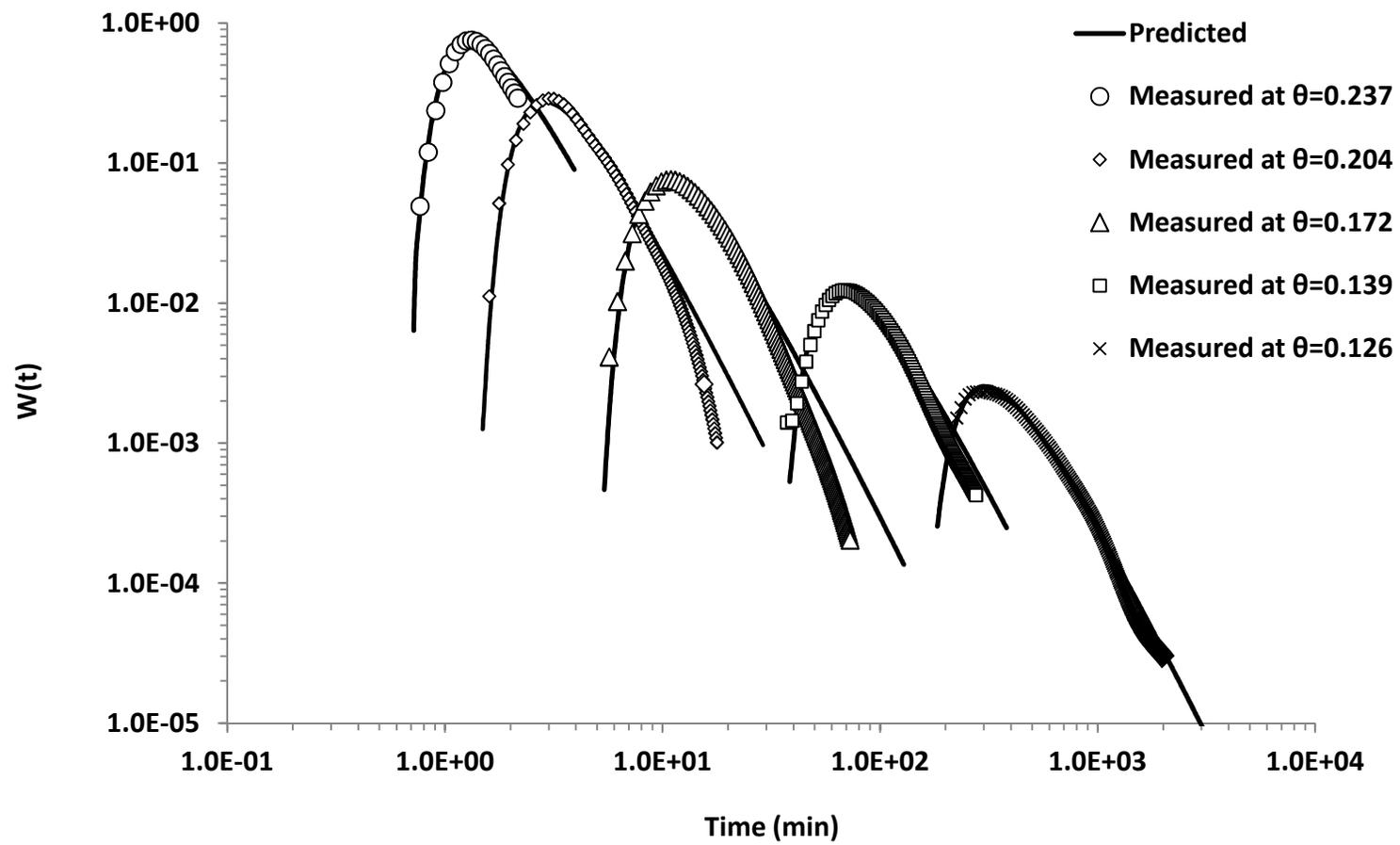

Figure 12

Table 1. The slope of arrival time distribution at large time for different model input parameters in 2-D porous media.

| Percolation universality class | D | $\phi$ | $\theta$ | $\theta_t$ | $\beta$ | power |
|---|---|---|---|---|---|---|
| Random | 1 | 0.6 | 0.6 | 0.2 | 1 | -1.626 |
| | 1 | 0.6 | 0.3 | 0.2 | 1 | -1.658 |
| | 1 | 0.6 | 0.6 | 0.1 | 1 | -1.641 |
| | 1 | 0.6 | 0.3 | 0.05 | 1 | -1.714 |
| | 1 | 0.6 | 0.6 | 0.2 | 0.63 | -1.901 |
| | 1 | 0.6 | 0.3 | 0.2 | 0.63 | -2.020 |
| | 2 | 0.6 | 0.6 | 0.2 | 1 | -1.444 |
| Invasion | 1 | 0.6 | 0.6 | 0.2 | 1 | -3.736 |
| | 1 | 0.6 | 0.3 | 0.2 | 1 | -3.724 |
| | 1 | 0.6 | 0.6 | 0.1 | 1 | -3.756 |
| | 1 | 0.6 | 0.3 | 0.05 | 1 | -3.983 |
| | 1 | 0.6 | 0.6 | 0.2 | 0.63 | -4.040 |
| | 1 | 0.6 | 0.3 | 0.2 | 0.63 | -3.910 |
| | 2 | 0.6 | 0.6 | 0.2 | 1 | -3.362 |

Table 2. The minimum and maximum slopes of arrival time distribution at large time for different model input parameters in 3-D porous media.

| class | | D=1 | D=2 | D=2.3 | D=2.6 | D=2.9 | D=2.95 |
|---|---|---|---|---|---|---|---|
| Random | Min | -2.118 | -2.043 | -2.099 | -1.848 | -1.815 | -1.848 |
| | Max | -2.015 | -1.923 | -1.961 | -1.453 | -1.510 | -1.630 |
| Invasion | Min | -3.115 | -3.152 | -3.179 | -3.189 | -3.185 | -3.218 |
| | Max | -2.958 | -2.907 | -2.801 | -2.469 | -2.415 | -2.385 |

Table 3. Physical properties for column displacement experiments (Jardine et al. [84]).

| Tracer | Tension head (cm) | water content (cm3/cm3) | Bulk density (gr/cm$^3$) | Sample length (cm) | Sample radius (cm) | Pore-water flux (cm/hr) |
|---|---|---|---|---|---|---|
| Bromide | 0 | 0.549 | 1.5 | 24 | 4.25 | 14.66 |
| Bromide | 10 | 0.533 | 1.5 | 24 | 4.25 | 2.82 |
| Bromide | 15 | 0.513 | 1.5 | 24 | 4.25 | 0.354 |

Table 4. Physical and hydraulic properties of Hanford sediment experiments reported by Cherrey et al. [86].

| Tracer | water content (cm3/cm3) | Sample length (cm) | Sample radius (cm) | Pore-water flux (cm/min) |
|---|---|---|---|---|
| Nitrate | 0.40 | 20 | 2.5 | 10.16 |
| Nitrate | 0.237 | 20 | 2.5 | 9.40 |
| Nitrate | 0.204 | 20 | 2.5 | 3.99 |
| Nitrate | 0.172 | 20 | 2.5 | 1.130 |
| Nitrate | 0.139 | 20 | 2.5 | 0.178 |
| Nitrate | 0.126 | 20 | 2.5 | 0.040 |